\documentclass[aps,pra,twocolumn,superscriptaddress]{revtex4-1}
\usepackage{amssymb,amsmath}
\usepackage[pdftex]{graphicx}
\usepackage[pdftex,colorlinks=true,bookmarks=true,citecolor=blue,linkcolor=blue,urlcolor=blue,breaklinks=true]{hyperref}
\newcommand{\bol}[1]{\boldsymbol{#1}}

\begin{document}
\title{Topological transition between competing orders in quantum spin chains}
\author{Shintaro Takayoshi}
\affiliation{Department of Quantum Matter Physics, University of Geneva,
Geneva 1211, Switzerland}
\affiliation{Max Planck Institute for the Physics of Complex Systems,
Dresden 01187, Germany}
\author{Shunsuke C. Furuya}
\affiliation{Condensed Matter Theory Laboratory, RIKEN,
Wako, Saitama 351-0198, Japan}
\author{Thierry Giamarchi}
\affiliation{Department of Quantum Matter Physics, University of Geneva,
Geneva 1211, Switzerland}

\date{\today}

\begin{abstract}
We study quantum phase transitions
between competing orders in one-dimensional spin systems.
We focus on systems that can be mapped to
a dual-field double sine-Gordon model
as a bosonized effective field theory.
This model contains two pinning potential terms of dual fields
that stabilize competing orders and allows different
types of quantum phase transition to happen
between two ordered phases.
At the transition point, elementary excitations
change from the topological soliton of
one of the dual fields to that of the other,
thus it can be characterized as a topological transition.
We compute the dynamical susceptibilities and the entanglement entropy, 
which gives us access to the central charge, 
of the system using a numerical technique of infinite time-evolving block decimation 
and characterize the universality class of the transition 
as well as the nature of the order in each phase. 
The possible realizations of such transitions in experimental systems
both for condensed matter and cold atomic gases are also discussed.
\end{abstract}

\maketitle

\section{Introduction}

Low dimensional quantum magnets show rich phase diagrams due to
the interplay between strong correlations and quantum fluctuations.
This competition is at the root of the existence of phases with very different physics,
separated by quantum phase transitions when parameters of the system are varied.
In one dimensional (1D) quantum magnets,
these transitions often have a topological nature.
The simplest example of such a transition is the one between
a massless phase dominated by XY correlations and the massive Ising phase existing
in an anisotropic Heisenberg spin-1/2 chain.
The universality class of this transition is
the celebrated Berezinskii-Kosterlitz-Thouless (BKT)
transition~\cite{Berezinskii1971JETP,Berezinskii1972JETP,Kosterlitz1973JPhysC}, 
which is characterized by a set of topological excitations.
A field theoretical description is instrumental in
understanding the properties of such transitions.
In the above mentioned case, the corresponding field theory
is the sine-Gordon model~\cite{Giamarchi2004Book}
and the low-energy excitations are solitons and carry a topological index.
Another example of system described by the sine-Gordon theory is
the Heisenberg chain with a staggered magnetic field such as
Cu benzoate~\cite{Oshikawa1997PRL,Affleck1999PRB,Dender1997PRL}.
A field theoretical approach to topological phases has been used with success
for more complicated phases, e.g. the Haldane phase in
$S=1$ quantum spin chains~\cite{Haldane1983PhysLettA,Haldane1983PRL}.

In this paper, we focus on the phase transitions in quantum magnets
which are caused by the competition between two dual fields having a topological nature.
Such systems are mapped onto a dual-field double sine-Gordon (DDSG)
model~\cite{Giamarchi1988JPhys,Lecheminant2002NPB,Delfino2002Inbook,Sarkar2016SciRep}.
This model contains two different potential terms pinning the dual fields.
If the strength of these potentials is varied,
the stabilized order is changed and a quantum phase transition occurs.
In addition to quantum magnets,
the DDSG model appears in a broad context such as in XY models
with symmetry breaking fields, in mixtures of electric charges
and magnetic monopoles~\cite{Jose1977PRB,Fertig2002PRL}, 
and in quantum ladder systems~\cite{Orignac2017PRB,Citro2018PRB,Robinson2018ArXiv}.
Experimentally the DDSG model has been realized
in the material $\mathrm{BaCo_{2}V_{2}O_{8}}$~\cite{Faure2018NatPhys}.
This compound has a strong Ising anisotropy and
when an external uniform magnetic field is applied,
an effective staggered field is introduced in the direction
perpendicular to both the anisotropy axis and the external magnetic field.
Thus the N\'eel orders along the anisotropy axis
and along the effective staggered field are competing in this system.
The quantum phase transition between them can be triggered
by increasing the strength of the external magnetic field,
and it is measured directly in inelastic neutrons scattering (INS) experiments.

In the following, we examine various possible realizations of
the DDSG model in quantum magnets,
and study quantitatively the resulting transitions.
We combine the field theory with a numerical analysis
based on the infinite time-evolving block decimation (iTEBD),
which utilizes a matrix product state (MPS) such as
the density matrix renormalization group~\cite{White1992PRL}.
We compute various observables such as the staggered magnetization,
the entanglement entropy and the dynamical spin-spin susceptibility.
In particular, the dynamical susceptibility not only
has a theoretical interest but also is directly related with the experiments
such as inelastic neutron scattering (INS),
electron spin resonance (ESR), and nuclear magnetic resonance (NMR).

This paper is organized as follows.
In Sec.~\ref{sec:bosonization}, we quickly review the bosonization
and give some examples of quantum spin systems
described by the DDSG model.
In Sec.~\ref{sec:QPT}, we study the quantum phase transition
between competing orders using the examples given
in Sec.~\ref{sec:bosonization}.
Section~\ref{sec:DynSuscep} discusses how the dynamical susceptibility
changes below and above the transition.
Section~\ref{sec:application} is devoted to discussing applications to
real materials. We summarize our results and discuss future problems
in Sec.~\ref{sec:summary}.

\section{Bosonization and dual-field double sine-Gordon model}
\label{sec:bosonization}

In this section, we briefly review the bosonization
of 1D spin chains~\cite{Giamarchi2004Book}.
We map the spin operators to bosonic scalar fields
using the formula,
\begin{equation}  \label{eq:bosonspin}
\begin{split}
 S_{j}^{z}&=-\frac{a}{\pi}\frac{d\phi(x)}{dx}
   +a_{1}(-1)^{j}\cos(2\phi(x))+\cdots,\\
 S_{j}^{+}&=e^{-i\theta(x)}
   [b_{0}(-1)^{j}+b_{1}\cos(2\phi(x))+\cdots],
\end{split}
\end{equation}
where $x=ja$ is a spatial coordinate ($a$ is the lattice constant) and
$a_{0}$, $b_{0}$ and $b_{1}$ are nonuniversal constants which can be estimated
numerically~\cite{Hikihara2004PRB,Takayoshi2010PRB,Bouillot2011PRB,Hikihara2017PRB}
$\phi(x)$ and $\theta(x)$ are dual bosonic fields
satisfying the commutation relation
$[\phi(x),\theta(x')]=-i\pi\vartheta_{\mathrm{step}}(x-x')$
($\vartheta_{\mathrm{step}}(x-x')$ is the step function).
The fields $2\phi(x)$ and $\theta(x)$ can be intuitively interpreted as
polar and azimuthal angles of the staggered magnetization.

The Hamiltonian of Heisenberg chains with an Ising anisotropy (XXZ models)
\begin{align}
 {\cal H}_{\mathrm{XXZ}}&=J\sum_{j}(S_{j}^{x}S_{j+1}^{x}
   +S_{j}^{y}S_{j+1}^{y}+\Delta S_{j}^{z}S_{j+1}^{z})
\label{eq:Hamil_XXZ}
\end{align}
is bosonized as
\begin{align}
 \mathcal{H}_{\mathrm{XXZ}}^{\mathrm{eff}}&=\frac{v}{2\pi}\int dx
   \Big[\frac{1}{K}\Big(\frac{d\phi(x)}{dx}\Big)^{2}
   +K\Big(\frac{d\theta(x)}{dx}\Big)^{2}\Big]\nonumber\\
   &\quad-\lambda\int dx\cos(4\phi(x))
   +\cdots,\nonumber
\end{align}
where $\lambda$ is some constant, $v$ is spinon velocity,
and $K$ is the Luttinger parameter.
The $\cos(4\phi(x))$ term has the scaling dimension $4K$,
and it is relevant in the Ising anisotropic ($\Delta>1$, $K<1/2$) region. 
It works as a potential to pin the field $\phi(x)$.
When $\phi(x)$ is fixed at $n\pi/2$ ($n$: integer),
the system has N\'eel order along the $z$ axis and the excitations are gapped.
If we add a pinning potential for $\theta(x)$,
it competes with the $\phi(x)$ pinning potential,
since $\phi(x)$ and $\theta(x)$ are conjugate they
cannot be simultaneously fixed. 
The resulting model is the DDSG model,
\begin{align}
 &\mathcal{H}_{\mathrm{DDSG}}=
   \frac{v}{2\pi}\int dx\Big[
   \frac{1}{K}\Big(\frac{d\phi(x)}{dx}\Big)^{2}
   +K\Big(\frac{d\theta(x)}{dx}\Big)^{2}\Big]\nonumber\\
   &-g_{1}\int dx\cos(m\phi(x))
    -g_{2}\int dx\cos(n\theta(x)).
\label{eq:DDSG}
\end{align}
where $m$ and $n$ are integers and
$g_{1}$, $g_{2}$ are nonuniversal constants.

In the following, we study several microscopic situations for
which the bosonized field theory is a DDSG model.

\subsection{XXZ model with a staggered magnetic field along the $x$ direction}
\label{sec:XXZhx}

Let us add a staggered magnetic field along the $x$ axis
$-h_{x}\sum_{j}(-1)^{j}S_{j}^{x}$
to the XXZ model~\eqref{eq:Hamil_XXZ}.
This staggered field is bosonized as
\begin{equation}
 -h_{x}\sum_{j}(-1)^{j}S_{j}^{x}
   =-h_{x}b_{0}\int dx\cos\theta(x)+\cdots.
\nonumber
\end{equation}
The $\cos\theta(x)$ term has a scaling dimension $1/(4K)$
and is relevant for $K>1/8$.
Therefore, the total bosonized Hamiltonian is
the DDSG model~\eqref{eq:DDSG} with $m=4$, $n=1$.
For $\Delta>1$ and $h_{x}=0$,
the ground state has N\'eel order (staggered magnetization)
along the $z$ axis and the $\phi$ field is pinned.
Since $\cos\theta(x)$ dominates over $\cos(4\phi(x))$
with increasing $h_{x}$ and the $\theta$ field is pinned,
there is a quantum phase transition.
The staggered field $h_{x}$ immediately creates a
finite staggered magnetization along the $x$ axis,
but the staggered magnetization along the $z$ axis
becomes $0$ in the high $h_{x}$ phase
and thus works as an order parameter.
Note that we could also use
$\langle \cos(\nu \theta(x))\rangle$ as an order parameter, 
where $\nu$ is any noninteger number (for example $\nu=1/2$)
since it becomes zero in the $\phi$ pinned phase
and nonzero only in the high field phase. 
Such order parameter is however nonlocal
in terms of the spin operators~\cite{Berg2008PRB} and 
thus its measurement can only be done in particular systems, 
as is discussed in Sec.~\ref{sec:application}. 
Using the spin current operator~\cite{Giamarchi2004Book}
\begin{equation}
 \mathcal{J}_{j}^{\mathrm{s}}\equiv
   \frac{i}{2}(S_{j}^{+}S_{j+1}^{-}-S_{j}^{-}S_{j+1}^{+})
   =-vK\frac{a}{\pi}\frac{d\theta(x)}{dx}+\cdots,
\nonumber
\end{equation}
$\cos(\nu \theta(x))$ is represented as
\begin{equation}
 \cos\Big(\frac{\nu\pi}{vK}\sum_{l=-\infty}^{j}\mathcal{J}_{l}^{\mathrm{s}}\Big)
   =\cos\Big(\nu\int_{-\infty}^{x}dx'\frac{d\theta(x')}{dx'}\Big)
     +\cdots.
\nonumber
\end{equation}
Thus nonlocal measurements
are needed for the experimental observation of
$\langle \cos(\nu \theta(x))\rangle$.
For quantities related to particle density (or $S^{z}$),
such nonlocal quantity could be measured in cold atomic systems 
(see Sec.~\ref{sec:coldatom}). 

Another order parameter which is local and can thus be directly measured in condensed matter experiments is the 
staggered magnetization $\cos(2\phi(x))$.
The lowest energy excitation is a soliton of
the $\phi(x)$ ($\theta(x)$) field in the low (high) $h_{x}$ phase.
The phase properties are summarized
in Table~\ref{tab:PhasePropertiesHx}.

\begin{table}[h]
\centering
\caption{Summary of the phase properties of the XXZ model
with a staggered magnetic field in the $x$ direction.}
\label{tab:PhasePropertiesHx}
\begin{tabular}{|l||l|l|} \hline
  & low $h_{x}$ phase & high $h_{x}$ phase \\ \hline
  pinned field & $\phi(x)$ & $\theta(x)$ \\
  $\langle\cos(2\phi(x))\rangle\propto\langle\sum_{j}(-1)^{j}S_{j}^{z}\rangle$ &
  $\neq 0$ & $0$ \\
  $\langle\cos\theta(x)\rangle\propto\langle\sum_{j}(-1)^{j}S_{j}^{x}\rangle$ &
  $\neq 0$ & $\neq 0$ \\
  $\langle\cos(\nu\theta(x))\rangle$ ($\nu$: noninteger) &
  $ 0$ & $\neq 0$ \\
  soliton &
  $\phi(x)=0\to\pi/2$ & $\theta(x)=0\to2\pi$ \\
  \hline
\end{tabular}
\end{table}

\subsection{XXZ model with XY anisotropy}

Let us now consider another type of perturbation
to the XXZ chain, which is the XY anisotropy.
When such a term is bosonized, it has the form of
\begin{equation}
 D_{xy}\sum_{j}(S_{j}^{x}S_{j+1}^{x}-S_{j}^{y}S_{j+1}^{y})
   =-D_{xy} c_{1}\int dx\cos(2\theta(x))+\cdots,
\nonumber
\end{equation}
where $c_{1}$ is a nonuniversal constant.
The $\cos(2\theta(x))$ term has the scaling dimension $1/K$
and it is relevant for $K>1/2$.
The total bosonized Hamiltonian is
the DDSG model~\eqref{eq:DDSG} with $m=4$, $n=2$, instead of $m=4$ and $n=1$ of the previous section.
In this case, the two cosine potential terms are
simultaneously marginal at $K=1/2$, and
a controlled perturbative renormalization can be
constructed~\cite{Giamarchi1988JPhys} around the marginal point.
The properties of such a transition will thus be quite different and are summarized
in Table~\ref{tab:PhasePropertiesAnis}.
\begin{table}[h]
\centering
\caption{Summary of the phase properties in the XXZ model
with XY anisotropy.}
\label{tab:PhasePropertiesAnis}
\begin{tabular}{|l||l|l|} \hline
  & low $D_{xy}$ phase & high $D_{xy}$ phase \\ \hline
  pinned field & $\phi(x)$ & $\theta(x)$ \\
  $\langle\cos(2\phi(x))\rangle\propto\langle\sum_{j}(-1)^{j}S_{j}^{z}\rangle$ &
  $\neq 0$ & $0$ \\
  $\langle\cos\theta(x)\rangle\propto\langle\sum_{j}(-1)^{j}S_{j}^{x}\rangle$ &
  $0$ & $\neq 0$ \\
  soliton &
  $\phi(x)=0\to\pi/2$ & $\theta(x)=0\to\pi$ \\
  \hline
\end{tabular}
\end{table}

\subsection{Other perturbations}

Although we focus mostly on the two above mentioned models below,
it is also possible to
consider other perturbations such as a staggered field along $z$ axis
$-h_{z}\sum_{j}(-1)^{j}S_{j}^{z}$ and
a dimerization $\delta\sum_{j}(-1)^{j}\bol{S}_{j}\cdot\bol{S}_{j+1}$.
These perturbations are bosonized as
\begin{align}
 -h_{z}\sum_{j}(-1)^{j}S_{j}^{z}
   =-h_{z}a_{1}\int dx\cos(2\phi(x))+\cdots,\nonumber\\
 \delta\sum_{j}(-1)^{j}\bol{S}_{j}\cdot\bol{S}_{j+1}
   =\delta d_{1}\int dx\sin(2\phi(x))+\cdots.\nonumber
\end{align}
These terms give another type of DDSG model,
but some of them can be related through a transformation since the fields $\phi$ and $\theta$ can be rescaled by the transformation 
\begin{equation}
\begin{split} 
 \phi &\to b \phi \\
 \theta &\to \frac1b \theta 
\end{split}
\end{equation}
that preserves the commutation relation. 
For example, the Heisenberg model with a staggered $z$ field
and XY anisotropy is equivalent to
the DDSG model~\eqref{eq:DDSG} with $m=2$, $n=2$.
This can be mapped to the $m=4$, $n=1$ case
through the transformation $\phi\to2\tilde{\phi}$,
$\theta\to\tilde{\theta}/2$, $K/4\to\tilde{K}$.
However the operators that correspond to local observable are different
since the formula~\eqref{eq:bosonspin} is unchanged.

\section{Quantum phase transition between competing orders}
\label{sec:QPT}

In this section, we study the properties of
the quantum phase transition between competing orders
for the models mentioned in Sec.~\ref{sec:bosonization}.

\begin{figure}[b]
\includegraphics[width=0.45\textwidth]{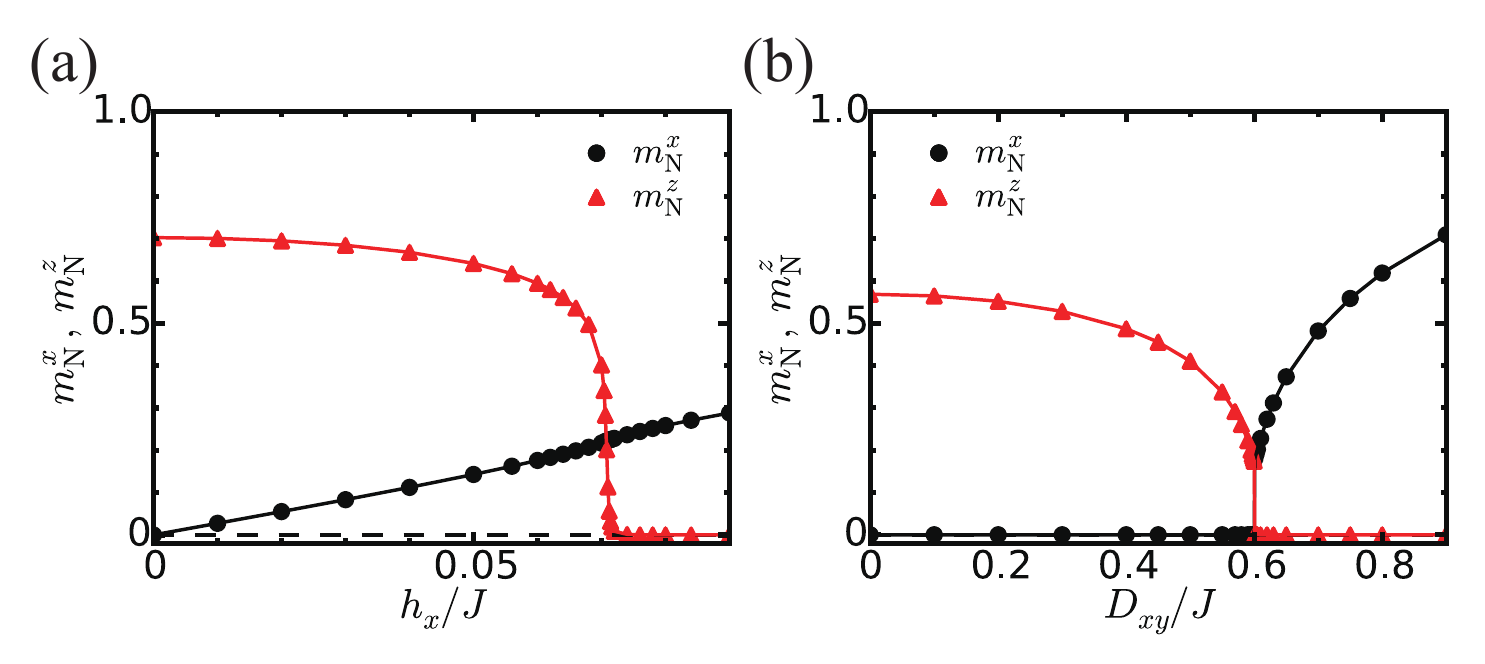}
\caption{Staggered magnetization curves
for $m_{\mathrm{N}}^{x}$ and $m_{\mathrm{N}}^{z}$
in the XXZ model with
(a) staggered $x$ field ($\Delta=1.9$)
and (b) XY anisotropy ($\Delta=1.6$).
The saturation value of $m_{\mathrm{N}}^{x(z)}$
is normalized to $1$.
}
\label{fig:magcur_Dxy_hx}
\end{figure}

\begin{figure}[t]
\includegraphics[width=0.45\textwidth]{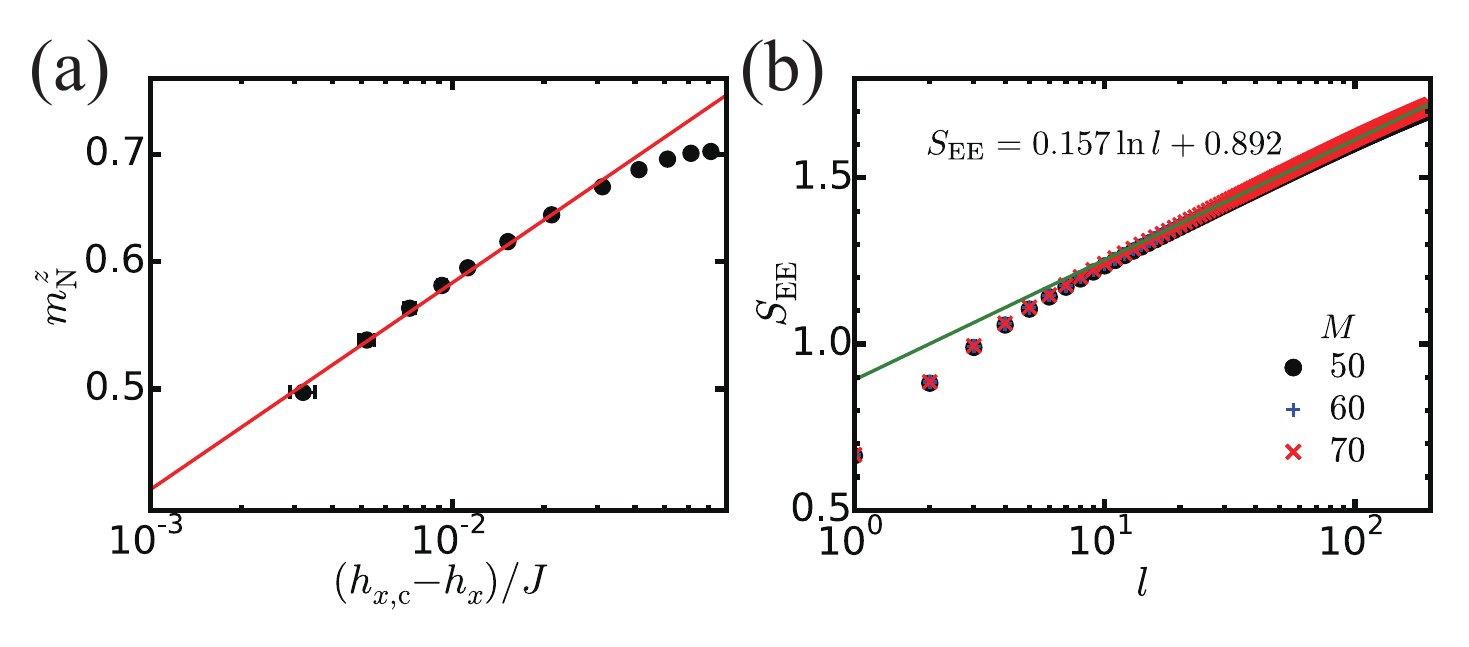}
\caption{(a) Log-log plot of $m_{\mathrm{N}}^{z}$ as a function of
$h_{x,\mathrm{c}}-h_{x}$.
(b) Semi-log plot of entanglement entropy for a finite interval
$S_{\mathrm{EE}}$ as a function of the size of the interval $l$
at $h_{x}=h_{x,\mathrm{c}}$.
$M$ is the bond dimension of MPS 
(see Appendix~\ref{sec:DetailNumerics}).
}
\label{fig:Ising}
\end{figure}

First, we consider the XXZ model with staggered $x$ field,
\begin{align}
 \mathcal{H}=\mathcal{H}_{\mathrm{XXZ}}
   -h_{x}\sum_{j}(-1)^{j}S_{j}^{x}.
\label{Hamil_XXZhx}
\end{align}
In Fig.~\ref{fig:magcur_Dxy_hx}(a),
we show the staggered magnetization per site
$m_{\mathrm{N}}^{x(z)}$ along $x(z)$ axis calculated by iTEBD.
The phase transition is characterized
by the disappearance of $m_{\mathrm{N}}^{z}$,
and the critical field is $h_{x,\mathrm{c}}/J\simeq0.071$.
Let us determine the universality class of this transition.
In Fig.~\ref{fig:Ising}(a), we show the log-log plot of
the order parameter $m_{\mathrm{N}}^{z}$
as a function of $h_{x,\mathrm{c}}-h_{x}$.
The fitting function is given as
$m_{\mathrm{N}}^{z}=1.055((h_{x,\mathrm{c}}-h_{x})/J)^{0.129}$,
and the critical exponent is $\beta=0.129\simeq 1/8$.
We also calculate the entanglement entropy for a finite interval.
When the system is bipartitioned into the subsystems $A$ and $B$,
where $A$ is an interval consisting of $l$ spins and $B$ is the remainder,
the reduced density matrix of the subsystem $A$ is defined as
$\rho_{A}=\mathrm{Tr}_{B}|\Psi\rangle\langle\Psi|$
($|\Psi\rangle$ is the ground state).
Then the entanglement entropy is represented as
$S_{\mathrm{EE}}=\mathrm{Tr}\rho_{A}\ln\rho_{A}$.
In systems described by a conformal field theory,
the entanglement scales as~\cite{Calabrese2004Jstat}
\begin{equation}
 S_{\mathrm{EE}}=\frac{c}{3}\ln l+\mathrm{const},
\label{eq:EntangleEnt}
\end{equation}
where $c$ is the central charge.
The entanglement entropy $S_{\mathrm{EE}}$ as a function of
the subsystem size $l$ that is calculated
at the transition point $h_{x,\mathrm{c}}$
is shown in Fig.~\ref{fig:Ising}(b).
When the data are fitted by~\eqref{eq:EntangleEnt},
the function is $S_{\mathrm{EE}}=0.157\ln l+0.892$
and the central charge is estimated as $c=0.471\simeq 1/2$.
These results $\beta\simeq 1/8$ and $c\simeq 1/2$ indicate that
the transition belongs to the Ising universality class.
In terms of a field theory, the DDSG model is equivalent to
two Majorana fermions~\cite{Lecheminant2002NPB,Tsvelik2012NJP}.
At the transition point, one of the Majorana fermions
is gapped out while the other remains gapless,
thus the transition is of the Ising type.

\begin{figure}[t]
\includegraphics[width=0.45\textwidth]{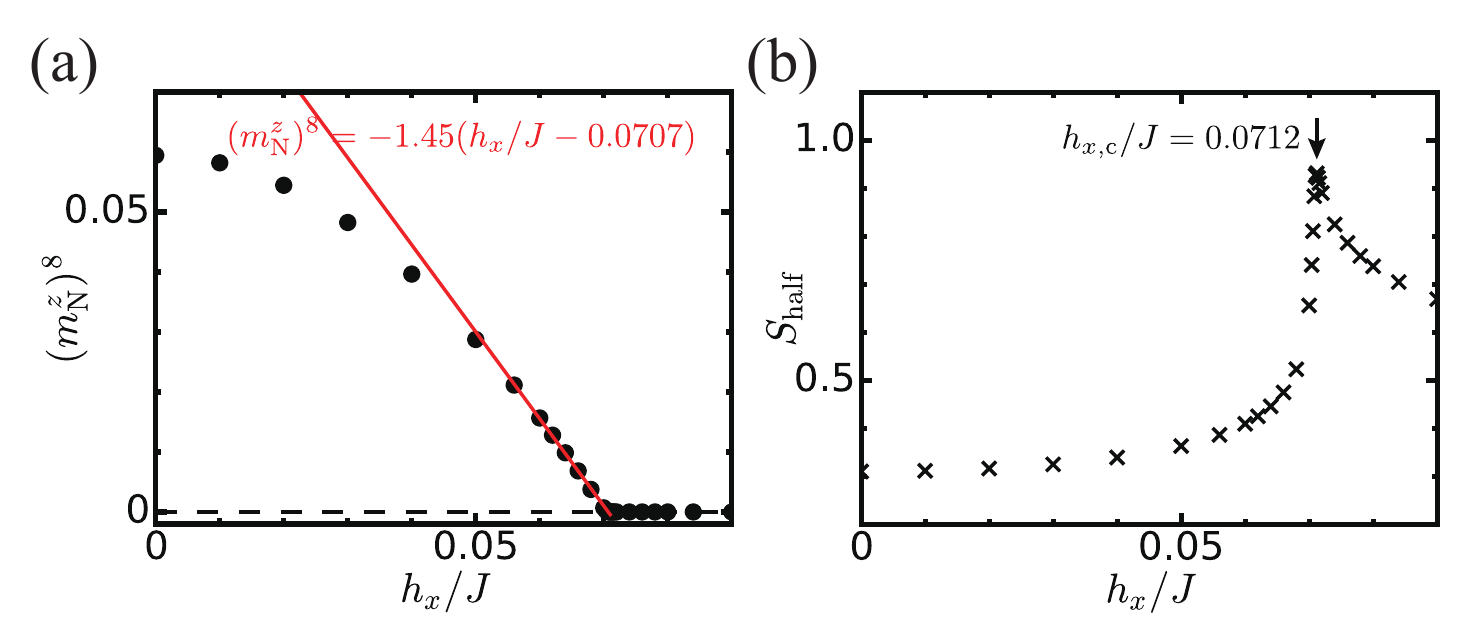}
\caption{Plot of (a) $(m_{\mathrm{N}}^{z})^{8}$ 
(b) half-infinite entanglement entropy $S_{\mathrm{half}}$ 
as a function of $h_{x}$. 
}
\label{fig:mz8_EE}
\end{figure}

In Fig.~\ref{fig:Ising}(a), 
we see that the data points are deviated from the fitting line 
in the region of 
and $(h_{x,\mathrm{c}}-h_{x})/J\gtrsim 0.03$. 
Let us comment on this point. 
Figure~\ref{fig:mz8_EE}(a) shows the plot of 
$(m_{\mathrm{N}}^{z})^{8}$ as a function of $h_{x}$. 
The solid line represents a linear fitting, 
and data points are away from the line in $h_{x}/J\leq 0.04$. 
This indicates that the deviation in the region of 
$(h_{x,\mathrm{c}}-h_{x})/J\gtrsim 0.03$ in Fig.~\ref{fig:Ising}(a) 
is due to getting away from the critical region. 
From the equation of the fitting line 
$(m_{\mathrm{N}}^{z})^{8}=-1.45(h_{x}/J-0.0707)$, 
the critical field is obtained as $h_{x,\mathrm{c}}/J=0.0707$. 
We can also determine $h_{x,\mathrm{c}}$ 
from the divergence of half-infinite entanglement entropy 
$S_{\mathrm{half}}$, which is calculated by 
the bipartition of the system into two half-infinite chains. 
In Fig.~\ref{fig:mz8_EE}(b), 
we plot the half-infinite entanglement entropy $S_{\mathrm{half}}$ 
as a function of $h_{x}$, 
and the critical value is $h_{x,\mathrm{c}}/J=0.0712$. 
Thus, it is estimated as $h_{x,\mathrm{c}}/J=0.071\pm 0.0003$, 
which causes the error bars in Fig.~\ref{fig:Ising}(a).

Next we consider the XXZ model with XY anisotropy,
\begin{align}
 \mathcal{H}=\mathcal{H}_{\mathrm{XXZ}}
   +D_{xy}\sum_{j}(S_{j}^{x}S_{j+1}^{x}-S_{j}^{y}S_{j+1}^{y}).
\label{Hamil_XXZDxy}
\end{align}
This Hamiltonian is nothing but the XYZ model,
which is exactly solvable~\cite{Baxter2016Book}.
Staggered magnetization $m_{\mathrm{N}}^{x}$
and $m_{\mathrm{N}}^{z}$ calculated by iTEBD is shown
in Fig.~\ref{fig:magcur_Dxy_hx}(b).
In contrast to Fig.~\ref{fig:magcur_Dxy_hx}(a),
the orders $m_{\mathrm{N}}^{x}$
and $m_{\mathrm{N}}^{z}$ are exclusively competing,
i.e., if one of the two orders is nonzero, the other is zero.
The critical value of $D_{xy}$ is $D_{xy,\mathrm{c}}=(\Delta-1)J$.
Since $J-D_{xy,\mathrm{c}}<J+D_{xy,\mathrm{c}}=\Delta J$,
the Hamiltonian is the easy-plane XXZ model at the critical point
and the ground state is Tomonaga-Luttinger liquid
(a conformal field theory with central charge $c=1$).
Hence the transition is the BKT type,
which is consistent with the renormalization analysis~\cite{Giamarchi1988JPhys}.

\section{Dynamical susceptibility}
\label{sec:DynSuscep}

Let us now compute how the critical behavior of the models
of Sec.~\ref{sec:QPT} can be measured experimentally. In addition to the static staggered magnetization,
we show that the dynamical susceptibility captures well the properties of the quantum phase transition.
This quantity is directly accessible in INS and ESR experiments.

The spin-spin retarded correlation function is defined as
\begin{equation}
 \chi^{\alpha\beta}(\bol{r},t)
   =-i\vartheta_{\mathrm{step}}(t)
     \langle[S_{\bol{r}}^{\alpha}(t),S_{0}^{\beta}(0)]\rangle,
\label{eq:RetCorr}
\end{equation}
where $\vartheta_{\mathrm{step}}(t)$ is the Heaviside function.
For 1D lattice systems, $\bol{r}$ is replaced with the site index $j$.
The dynamical susceptibility is obtained from the Fourier transform
of the retarded correlation function~\eqref{eq:RetCorr},
\begin{equation}
 \chi^{\alpha\beta}(\bol{q},\omega)
   =\int_{-\infty}^{\infty}dt\sum_{\bol{r}}
     e^{i(\omega t-\bol{q}\cdot\bol{r})}
     \chi^{\alpha\beta}(\bol{r},t)
\label{eq:SuscepFourier3D}
\end{equation}
This quantity is related to
the differential scattering cross section of INS by
\begin{align}
  \frac{d^{2}\sigma}{d\Omega dE}\propto
   \frac{|\bol{q}_{\mathrm{out}}|}{|\bol{q}_{\mathrm{in}}|}|F(\bol{Q})|^{2}
   \sum_{\alpha,\beta=x,y,z}&
   \Big(\delta_{\alpha\beta}
     -\frac{Q_{\alpha}Q_{\beta}}{|\bol{Q}|^{2}}\Big)\nonumber\\
   &\times
   \mathrm{Im}\chi^{\alpha\beta}(\bol{Q},\omega),
\label{eq:CrossSec}
\end{align}
where $F(\bol{Q})$ is the magnetic form factor
and $\bol{q}_{\mathrm{in}}$, $\bol{q}_{\mathrm{out}}$ is the direction of
incoming and outgoing fluxes, respectively.
$\bol{Q}$ is a scattering vector
defined as $\bol{Q}=\bol{q}_{\mathrm{in}}-\bol{q}_{\mathrm{out}}$.
If the system is $U(1)$ symmetric
(i.e., $\sum_{j}S_{j}^{z}$ is conserved),
Eq.~\eqref{eq:CrossSec} is rewritten as~\cite{Lovesey1986Book}
\begin{align}
  \frac{d^{2}\sigma}{d\Omega dE}\propto&
   \frac{|\bol{q}_{\mathrm{out}}|}{|\bol{q}_{\mathrm{in}}|}|F(\bol{Q})|^{2}
   \Big\{\Big(1-\frac{Q_{z}^{2}}{|\bol{Q}|^{2}}\Big)
     \mathrm{Im}\chi^{zz}(\bol{Q},\omega)\nonumber\\
   &\qquad
   +\Big(1+\frac{Q_{z}^{2}}{|\bol{Q}|^{2}}\Big)
     \mathrm{Im}\chi^{xx}(\bol{Q},\omega)\Big\},
\label{eq:CrossSec2}
\end{align}
since $\chi^{xx}=\chi^{yy}$.
In ESR experiments, since electromagnetic waves
in the GHz frequency region are used,
the wavelength is much larger than the lattice constant
and only the response at $|\bol{q}|= 0$ is relevant.
When such electromagnetic waves are applied to the system,
the energy absorption rate is given by
\begin{equation}
 I(\omega)\propto \omega\mathrm{Im}
   \chi^{\alpha\alpha}(\bol{q}=0,\omega),
\end{equation}
where $\alpha$ is the direction of oscillating magnetic field.
$I(\omega)$ corresponds with spectrum of ESR.

We compute the dynamical susceptibility numerically.
We first obtain the ground state of the system
by infinite density matrix renormalization group (iDMRG)~\cite{Mcculloch2008arXiv},
then perform the time evolution by iTEBD~\cite{Vidal2007PRL}
with the infinite boundary condition~\cite{Phien2012PRB}.
In this way, we can calculate space-time correlation function
$\langle S_{\bol{r}}^{\alpha}(t)S_{0}^{\beta}(0)\rangle$,
and dynamical susceptibility through Fourier transform.
The details of numerical calculation are given
in Appendix~\ref{sec:DetailNumerics}.

\begin{figure}[t]
\includegraphics[width=0.45\textwidth]{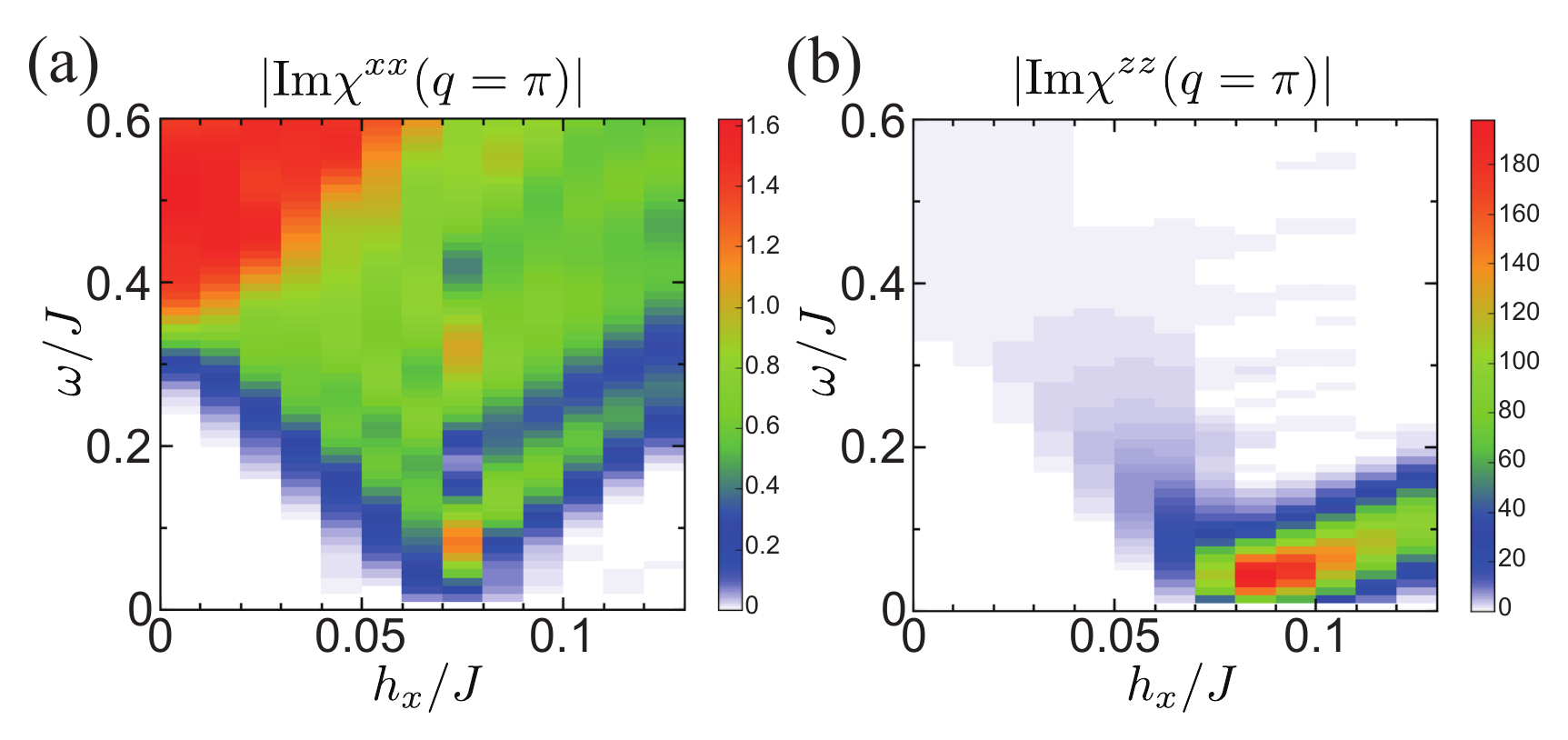}
\caption{Dynamical susceptibility
(a) $\chi^{xx}(q=\pi)$ and (b) $\chi^{zz}(q=\pi)$
for the XXZ model ($\Delta=1.9$) with staggered $x$ field.
The dominant low energy excitation in the low (high) $h_{x}$ phase
corresponds to $\chi^{xx}$ ($\chi^{zz}$).
We see that $\chi^{zz}$ diverges at the transition point
$h_{x}/J\simeq 0.071$ while $\chi^{xx}$ does not.
}
\label{fig:chiqw_hx}
\end{figure}

In Fig.~\ref{fig:chiqw_hx},
we show the dynamical susceptibility at $q=\pi$ in the XXZ model
with staggered $x$ field~\eqref{Hamil_XXZhx}.
In the low (high) $h_{x}$ phase,
the dominant low energy elementary excitation corresponds to
$\chi^{xx}$ ($\chi^{zz}$).
The order is in the $z$ direction at $h_{x}=0$,
and $m_{\mathrm{N}}^{z}$ decreases while $m_{\mathrm{N}}^{x}$ increases
as $h_{x}$ becomes larger.
Above the critical $h_{x}$, the order is in the $x$ direction.
Hence the behavior of $\chi^{xx}$ and $\chi^{zz}$ indicates that
the low energy excitation is generated by a spin flip.
We can also see that $\chi^{zz}$ diverges at the transition point
while $\chi^{xx}$ does not in Fig.~\ref{fig:chiqw_hx}.
That is because $m_{\mathrm{N}}^{z}$ becomes zero at the transition point
while $m_{\mathrm{N}}^{x}$ changes smoothly.
[see Fig.~\ref{fig:magcur_Dxy_hx}(a)].

\begin{figure}[t]
\includegraphics[width=0.45\textwidth]{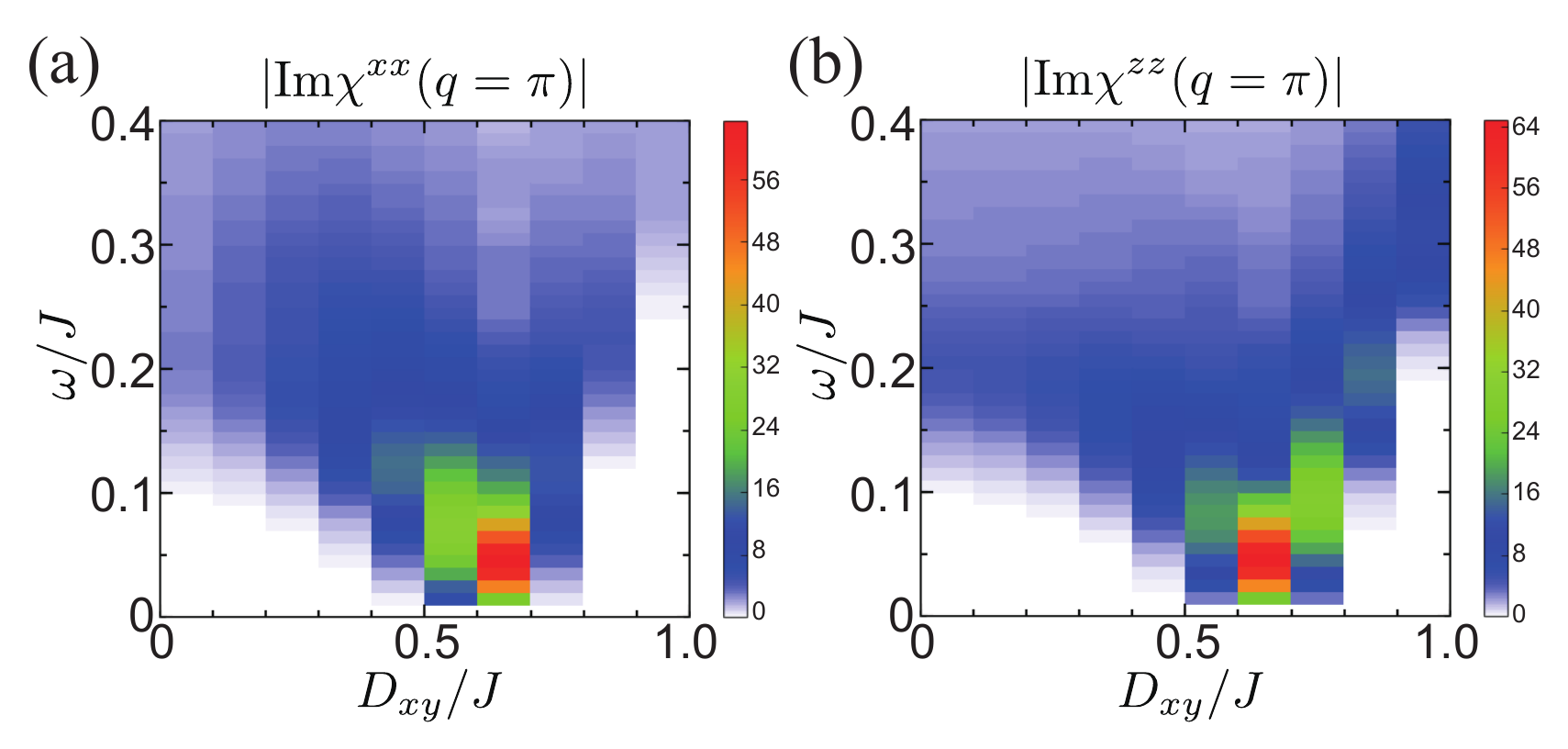}
\caption{Dynamical susceptibility
(a) $\chi^{xx}(q=\pi)$ and (b) $\chi^{zz}(q=\pi)$
for the XXZ model ($\Delta=1.6$) with XY anisotropy.
Both $\chi^{xx}$ and $\chi^{zz}$ diverge
at the transition point $D_{xy}/J=0.6$.
}
\label{fig:chiqw_Dxy}
\end{figure}

Let us now compare with the dynamical susceptibility at $q=\pi$ for the XXZ model
with XY anisotropy~\eqref{Hamil_XXZDxy} in Fig.~\ref{fig:chiqw_Dxy}.
Similarly to the staggered $x$ field case,
in the low (high) $D_{xy}$ phase,
the dominant elementary excitation corresponds to $\chi^{xx}$ ($\chi^{zz}$).
There are however an important difference on the susceptibilities,
which stems from the different nature of the transition.
It is directly visible that both $\chi^{xx}$ and $\chi^{zz}$
diverge at the transition point in Fig.~\ref{fig:chiqw_Dxy}.
This is the consequence of the exclusive competition
between $m_{\mathrm{N}}^{x}$ and $m_{\mathrm{N}}^{z}$,
both of which become zero at the transition point
[see Fig.~\ref{fig:magcur_Dxy_hx}(b)].

\begin{figure}[t]
\includegraphics[width=0.45\textwidth]{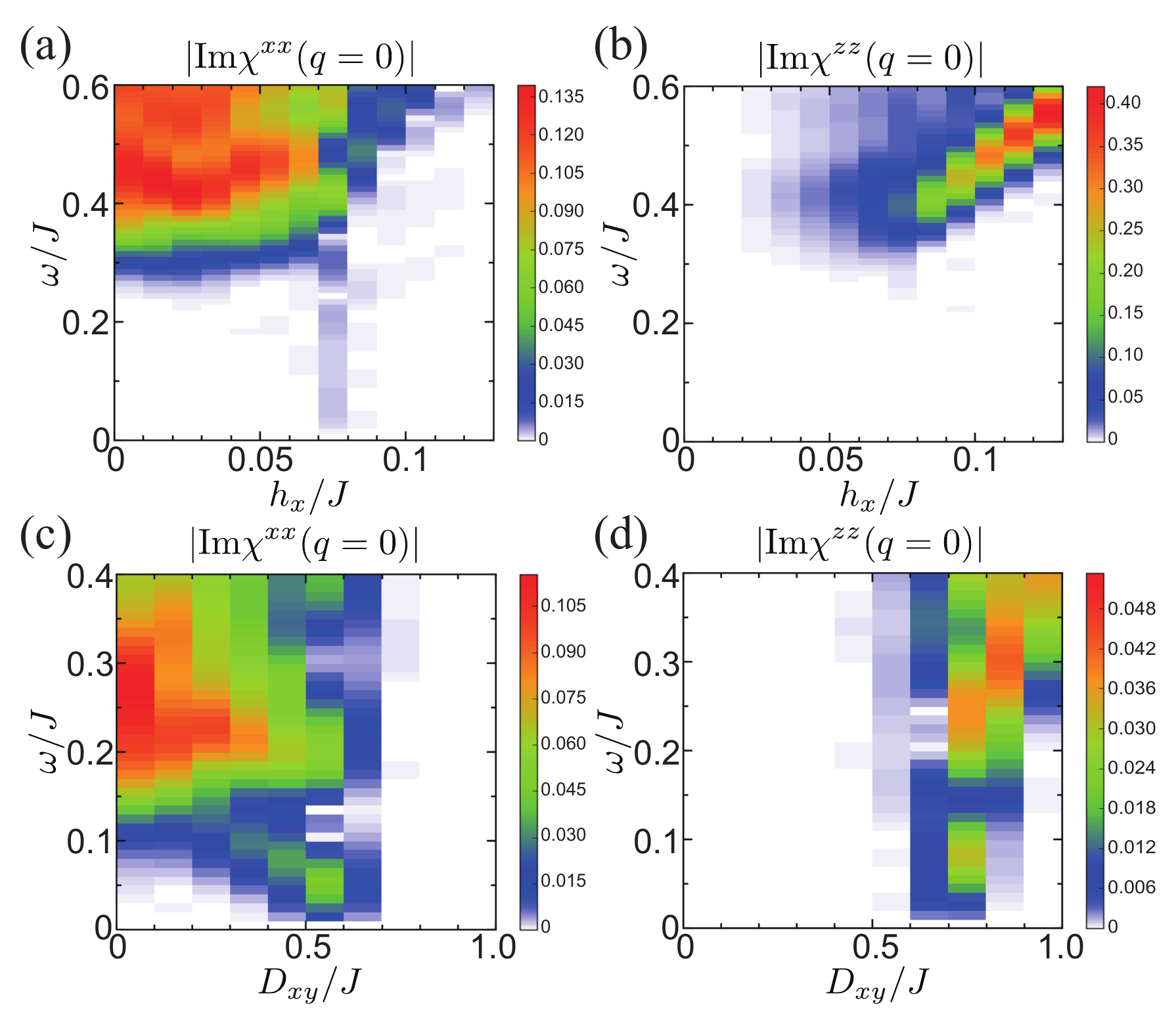}
\caption{
Dynamical susceptibility
(a) $\chi^{xx}(q=0)$ and (b) $\chi^{zz}(q=0)$
for the XXZ model ($\Delta=1.9$) with staggered $x$ field 
and (c) $\chi^{xx}(q=0)$ and (d) $\chi^{zz}(q=0)$
for the XXZ model ($\Delta=1.6$) with XY anisotropy.
}
\label{fig:chiqw_q0}
\end{figure}

We also discuss the dynamical susceptibility at $q=0$ 
which is relevant with ESR experiments.
Figure~\ref{fig:chiqw_q0} shows $\chi^{xx}(q=0)$ and $\chi^{zz}(q=0)$ 
for the XXZ model ($\Delta=1.9$) with staggered $x$ field 
and with XY anisotropy.
We first note that the intensity of the dynamical susceptibility 
is extremely small at $q=0$ compared with $q=\pi$ 
since antiferromagnetic correlation is dominant in the present system. 
As seen in Figs.~\ref{fig:chiqw_q0}(a) and (b),
gap does not close at $q=0$ 
for the XXZ model with staggered $x$ field. 
Small intensity of the low energy region ($\omega/J\lesssim 0.3$) 
near the critical field $h_{x}\simeq 0.07$ is numerical artifact. 
On the contrary, Figs.~\ref{fig:chiqw_q0}(c) and (d) show that 
gap closes at $q=0$ for the XXZ model with XY anisotropy. 
This is natural since the critical point corresponds to 
an easy plain XXZ model and 
the gapless des Cloizeaux-Pearson mode exists at $q=0$. 

As for the XXZ model with staggered $x$ field, 
the band at $q=\pi$ is folded to the band at $q=0$ 
due to the perturbation that breaks one-site translational symmetry. 
Thus, ESR measurements captures the mixing 
of $q=0$ and $q=\pi$ components of dynamical susceptibility. 
This effect is seen in Cu benzoate~\cite{Oshikawa2002PRB}, 
KCuGaF$_{6}$~\cite{Furuya2012PRL}, 
and $\mathrm{BaCo_{2}V_{2}O_{8}}$~\cite{Kimura2007PRL}.
The similar mixing is also measured 
in $\mathrm{(C_{7}H_{10}N)_{2}CuBr_{4}}$~\cite{Ozerov2015PRB}. 

The above calculations clarifies that the spin-spin susceptibility
shows very clear signatures of the nature of these two different
topological transitions. Although these measurements do not directly
give access to the nonlocal (topological) order, they nevertheless
provide clear signatures of the change of the nature of the excitations.

\section{Application to real materials}
\label{sec:application}

In the above, we discussed the models that can be mapped to
DDSG models and their quantum phase transitions.
In order to apply the above theoretical analysis to realistic materials,
one has to consider several important elements depending on
whether the system is condensed matter or cold atomic gas.

\subsection{Condensed matter systems}

For the condensed matter realizations,
two elements are to be taken into account.
First, in the present experiments, one can expect to measure
only the local observable (magnetization, spin-spin susceptibility, etc.).
Nonlocal order parameters
(e.g., $\cos(\theta(x)/2)$ in Sec.~\ref{sec:XXZhx})
are difficult to measure experimentally in condensed matter systems.
Second, in quasi-1D materials, spin chains are coupled
and form three dimensional system
while the analysis done in the previous parts is strictly 1D.

Recently, the DDSG model discussed above was found to be realized
in the compound $\mathrm{BaCo_{2}V_{2}O_{8}}$~\cite{Faure2018NatPhys}.
In this material, Co$^{2+}$ ions effectively form
the $S=1/2$ quasi-1D antiferromagnet with Ising anisotropy.
When an external magnetic field
perpendicular to the anisotropy axis is applied in this system,
an effective staggered transverse field arises
since nondiagonal components of $g$ tensor are nonzero due
to the slight deviation of the magnetic principal axes
from the crystallographic axes~\cite{Kimura2013JPSJ}.
The model Hamiltonian of this compound is essentially
equivalent to the XXZ model with staggered $x$ field~\eqref{Hamil_XXZhx},
and the quantum phase transition discussed
in Sec.~\ref{sec:XXZhx} happens.
Note that an effective staggered field 
$-h_{\mathrm{eff}}\sum_{j}(-1)^{j}S_{j}^{z}$ along the $z$ axis
arises from the interchain interaction, determined self-consistently, 
with the N\'eel order along the $z$ axis in the mean field theory 
has also to be taken into account \cite{Faure2018NatPhys}. 
Due to this staggered $z$ field,
the critical field is shifted to a higher value
than the case without the interchain interaction
and the gap opens at the transition point with $h_{\mathrm{eff}}=0$.
Thus, the gap is not closed at the quantum phase transition
caused by the transverse field in $\mathrm{BaCo_{2}V_{2}O_{8}}$.
As discussed in Sec.~\ref{sec:DynSuscep},
the dynamical susceptibility is measured by INS experiments.
For a direct comparison with the neutrons,
one has to use the actual position of the spin sites (the Co$^{2+}$ ions)
in the Fourier transform of retarded correlation function
since the neutrons are directly sensitive to the actual position of the spins.

It would be interesting if other examples of the topological transitions 
discussed in the previous sections also could be realized. 
The potential of the field $\phi$ is provided by dimerization, 
Ising anisotropy, and staggered Dzyaloshinskii-Moriya (DM) interaction 
$\sum_{j}(-1)^{j}\bol{D}\cdot(\bol{S}_{j}\times\bol{S}_{j+1})$ 
with $D\parallel z$ axis. 
The strategy for material search is to find systems that 
have these perturbations as well as nondiagonal staggered $g$ tensor. 
The application of effective staggered field introduces 
effective staggered field, which gives the potential of the field $\theta$. 
Then the transition is provoked by increasing the external field. 
In addition to spin chains, searching for materials which realize 
the DDSG model in spin ladders with magnetic anisotropy 
or DM interaction is an interesting future direction.

\subsection{Cold atomic systems}
\label{sec:coldatom}

Another important route to realize the topological transitions 
described in the previous sections is provided 
by cold atomic systems~\cite{Bloch2008RMP,Ritsch2013RMP}. 
Although initial simulations of quantum magnetism were done 
in bosonic systems by using the mapping 
between spin-1/2 and hard core bosons~\cite{Struck2011Science,Vedmedenko2013NJP} 
and thus the realization is limited to XX models 
due to the absence of long range interactions, 
recent advance allows to probe the quantum magnetism in fermionic systems as well. 
Short-range quantum magnetism has been observed 
for ultracold fermions in an optical lattice~\cite{Greif2013Science},
and measurements of various physical quantities such as
dynamical structure factor~\cite{Landig2015NatComm} and magnetic order and 
correlations~\cite{Parsons2016Science,Boll2016Science,Cheuk2016Science}. 
In addition to systems with fermions, 
quantum simulation of spin systems are also realized
by using Rydberg atoms~\cite{Browaeys2016JPhysB,Lienhard2018PRX}.

There are several advantages for the cold atomic realization.
The first is the controllability of parameters.
While the parameters are fixed for each material
in condensed matter systems, particle-particle interaction
can be varied by using Feshbach resonance in cold atomic systems.
Controlling the population of up-spins and down-spins 
allows the equivalent of a magnetic field along $z$. 
The second advantage is that cold atomic systems provide
the probes complementary to the condensed matter ones, 
in particular to measure nonlocal order parameters. 
For example, a string order parameter in the Haldane phase
can be observed by repeating
snapshot measurements~\cite{Endres2011Science}
in cold atomic systems.
This technique can be also potentially applicable for measuring
nonlocal order parameters such as $\cos(\theta(x)/2)$
discussed in Sec.~\ref{sec:XXZhx}. 
Measurements are so far limited to equal time correlations 
but schemes have been proposed to overcome such limitations~\cite{Knap2013PRL}. 

One of the challenges in this field is
cooling the system enough to simulate the low temperature phenomena
of the corresponding condensed matter systems.
However, since the experimental technique of cooling has been
improving~\cite{Mazurenko2017Nature},
we can expect that some of the phases described here 
could be observed in the near future.

\section{Conclusion}
\label{sec:summary}

We studied quantum phase transitions between competing orders
in the models which is mapped to
the DDSG field theory.
We specifically considered two types of systems:
the XXZ chain with staggered $x$ field and
with XY anisotropy.
The universality class of the transition is
of the Ising type in the former case
while it is of the BKT type in the latter case.
We showed numerically that
the difference of the transition properties
appears in the dynamical susceptibilities,
which can be directly compared with the spectra measured
by INS experiments.
We discussed the possibility of observation of the phases and the phase transitions studied 
in the present paper in condensed matter systems and cold atomic ones. 
For condensed matter realizations, one of the quantum phase transition 
between competing orders has been seen in a real material $\mathrm{BaCo_{2}V_{2}O_{8}}$,
which is a quasi-1D Heisenberg antiferromagnet with Ising anisotropy~\cite{Faure2018NatPhys}. 
Other quantum spin systems either chains or ladders 
with anisotropic perturbations could serve as a basis 
for studying the other universality classes discussed here. 
In that respect the dynamical susceptibilities, directly measured 
by INS or ESR experiments, computed in the present paper, 
provide a clear distinction between the various transitions and 
can thus be used as an experimental signature. 

Another broad class of systems in which the phenomena can be investigated 
is provided by cold atomic systems of fermions or Rydberg atoms. 
Such systems have the advantage of a good control 
of the various parameters in the Hamiltonian 
as well as the possibility of measure the nonlocal (topological) order parameters 
which are a direct signature of the various phases. 
Relatively high temperature as well as the size limitation 
is the current drawbacks, but the situation is rapidly evolving. 
These systems also offer the fascinating possibility to study time-dependent Hamiltonians, 
allowing to investigate the effect of time dependent perturbations in the future, 
either quenches or periodic perturbations (Floquet systems) 
on such topological phase transitions. 

\acknowledgements

The authors gratefully thank the many fruitful discussions with
Q. Faure, B. Grenier, S. Petit, V. Simonet and Ch. R\"uegg on quantum spin chains. 
S. T. is supported by the Swiss National Science Foundation
under Division II and ImPact project (No. 2015-PM12-05-01)
from the Japan Science and Technology Agency.
S. C. F. is supported by JSPS KAKENHI (No. JP16J04731).

\appendix
\section{Details of numerical simulations}
\label{sec:DetailNumerics}

In this appendix, we describe the detail of numerical simulations.
Time evolution is calculated by iTEBD~\cite{Vidal2007PRL}
after the ground state is obtained by iDMRG~\cite{Mcculloch2008arXiv}.
The iTEBD uses the MPS representation of quantum states,
and the time evolving operator
is applied through the second order Trotter decomposition.
Time is discretized with the unit $dt/J^{-1}=0.05$ in this study.
The initial state (ground state) is represented as infinite MPS,
which assumes translational invariance of the system,
but in order to calculate the space-time correlation function,
we have to break the translational invariance
by applying an operator at $t=0,j=0$.
Thus, we prepare a finite spatial interval
and the matrices at both edges of the interval is determined
in the way that they represent a semi-infinite extension
of the system, which is called
the infinite boundary condition~\cite{Phien2012PRB}.
The advantage of this method is that there is no finite-size effect.
The space-time correlation function Eq.~\eqref{eq:RetCorr}
is calculated for a finite temporal interval $0\leq t \leq T$,
and dynamical susceptibility is obtained
as the numerical Fourier transform
of the space-time correlation function.
Gaussian filter is utilized in the Fourier transformation,
\begin{equation}
 \chi(q,\omega)
   =\int_{-T}^{T}dt\sum_{r}e^{i(\omega t-qr)}\chi(r,t)G(t),
\nonumber
\end{equation}
where $G(t)=e^{-(2t/T)^{2}}$.

\begin{figure}[b]
\includegraphics[width=0.45\textwidth]{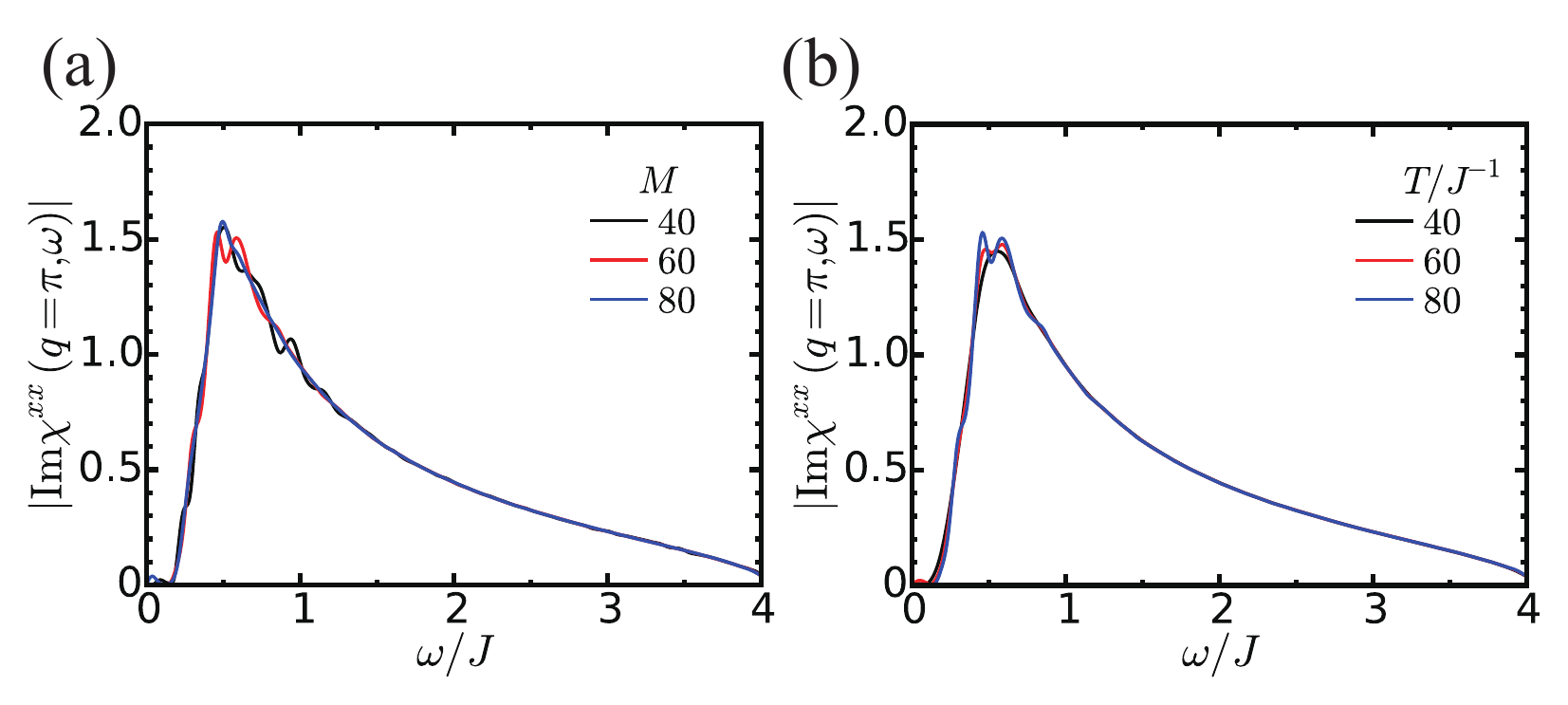}
\caption{The dependence of iTEBD calculations
(a) on the truncation dimension $M$ with fixed $T/J^{-1}=80$
and (b) on the temporal interval $T$ with fixed $M=60$.
The results of $\mathrm{Im}\chi^{xx}(q=\pi,\omega)$
for the model~\eqref{Hamil_XXZhx} are shown
with $\Delta=1.9$ and $h_{x}/J=0.02$.
}
\label{fig:M_T_depend}
\end{figure}

In the iTEBD and iDMRG calculations,
quantum states are optimally approximated by MPS with finite bond dimension
(also called truncation dimension) $M$.
As the bond dimension $M$ is larger, the calculation is more precise.
In Fig.~\ref{fig:M_T_depend}(a),
we show $\chi^{xx}(q=\pi,\omega)$ calculated with Eq.~\eqref{Hamil_XXZhx}
for different bond dimensions $M=40,60,80$ while $T/J^{-1}=80$ is fixed.
We can see that the dependence of the result on $M$ is small.
In the real-time calculation, an error also arises
from a finite time effect.
Figure~\ref{fig:M_T_depend}(b) shows $\chi^{xx}(q=\pi,\omega)$
calculated with Eq.~\eqref{Hamil_XXZhx}
for final time $T/J^{-1}=40,60,80$ while $M=60$ is fixed.
The dependence of the result on $T$ is also small.

\end{document}